# Application of Global Route-Planning Algorithms with Geodesy

William C. da Rosa, Iury V. de Bessa, Lucas C. Cordeiro
Federal University of Amazonas, Brazil

**Abstract**

**Global Route-Planning Algorithms (GRPA) are required to compute paths between several points located on Earth's surface. A geodesic algorithm is employed as an auxiliary tool, increasing the precision of distance calculations. This work presents a novel simulator for three GRPA – A\*, LPA\* and D\*Lite – implemented to solve the shortest path problem for points located at different cities. The performance of each algorithm is investigated with a set of experiments, which are executed to check the answers provided by the algorithms and to compare their execution time. It is shown that GRPA implementations with consistent heuristics lead to optimal paths. The noticeable differences among those algorithms are related to the execution time after successive route calculations.**

## I. INTRODUCTION

Several Global Route-Planning Algorithms (GRPA), such as A* [1], LPA* [2] and D*Lite [3], have been developed in the last decades to offer paths with the lowest cost in short time intervals. These algorithms work with a discrete representation of an environment (i.e., route graph) [4], which describes different paths in which an agent may use for navigation. GRPA may also be referred to as planners.

A graph is composed by vertices and edges, where an edge connects two consecutive vertices directly. The route is built after successive connections between vertices.

In this paper, vertices are considered to be points located at different positions on Earth's surface. They are connected by edges, which have costs associated to them. These costs are calculated as the length of geodesics on the ellipsoid, which are curves on the ellipsoid's surface. Thus, the values of edge costs must be calculated with a high precision to guarantee the computation of optimal paths. For this purpose, a geodesic algorithm is applied.

The geodesic algorithm adopted here is Karney's Geodesic Algorithm [5], [6]. An implementation of that algorithm is proposed for Interactive Data Language (IDL) [7]. It details the canonical representation of two vertices linked by an edge, located on Earth's surface, and defines numerical intervals to determine whether the pair of points is antipodal, i.e., it checks whether such vertices are located at diametrically opposite positions on a sphere or ellipsoid surface.

In order to evaluate the planners performance and to calculate geodesic lengths, simulators are usually employed. Two important simulators may be mentioned, one for geodesic calculations and another one for a planner. The online simulator for geodesics, named as "Online geodesic calculations using the GeodSolve utility" [8], solves direct and inverse geodesic problems, offering tools to setup angular outputs, numerical precision and geodesic lengths for different ellipsoids. The simulator for a planner behavior, named as *D\*Lite Demonstration* [9], is an open-source simulator for Java platform. It has a friendly interface, enabling the user to edit grid maps and execute them in debug mode. It is a practical tool for the comprehension of this planner, since it demonstrates how the variables, necessary for route computing, evolve during the algorithm execution.

Both simulators are useful to validate different types of calculations. The first simulator provides a reliable answer for the geodesic length between two points on Earth's surface. The second one applies a specific planner to quickly compute a path in a graph with adjustable edge costs. Thus, it validates the correct performance of such planner in a graph with limited size, where all vertices have the same degree and are distributed in a geometric standard, i.e., a square-grid board.

It is also useful to join and expand the functionalities of both simulators, creating a discrete environment, where vertices are represented in real world locations and have different direct connections from each other. A simulator that combines pathfinding and geodesic abilities computes optimal paths on Earth surface, in a graph where vertices do not follow a geometric standard and have different degrees. It generalizes the pathfinding problem to real situations.

A simulator was created in this work to employ three different planners – A*, LPA*, and D*Lite – which must use geodesic functions to compute optimal paths on Earth's surface. The performances of planners are also compared in terms of execution time.

Our simulator, named Geodesic Path Comparer (GPC), carries out an implementation of Karney's Algorithm. GPC considers a simplification made in [7], which determines the antipodal status of two points based on their geographic coordinates. Then, the geodesic lengths calculated by GPC are validated by comparison to the ones provided by [8]. GPC also expands concepts illustrated in [9] by generating paths with vertices whose placements follow no geometric standard.

The main contributions of this article are: the employment of geodesy and adaptive data structures for the creation of GPC; and the evaluation of planners' behaviors in graphs with nodes at irregular positions and with different degrees.

## II. PRELIMINARIES

### 2.1. Path Planning Algorithms

A* [1] is a best-first search algorithm, which uses heuristics and traversal costs, obtained from expanded nodes, to maintain a priority queue that represents a set of nodes of the graph. The order of nodes expansions follows the queue order, making the search for a path faster if compared to other algorithms with no heuristics or previous information from the graph.

LPA* [2] is an expansion of A*, which is able to keep information between successive searches, provided that starting and ending vertices of a route remain static. LPA* does not need to compute all information for the graph from the scratch. It is possible by creating an extra variable for traversal costs and by checking nodes consistency, using heuristics to detect relevant nodes for new route computation.

D*Lite [3] is similar to LPA*, but it is more suitable for graphs where a starting node changes over time. The search direction is the inverse of the adopted by LPA*, which can exempt nodes already traveled in the graph. D*Lite is able to update its priority queue without constant reordering, which is important for a fast and optimized response. This feature is derived from D* [10], an algorithm whose behavior is similar to A*, with the exception that edges costs are variable. In this paper, all heuristics are consistent, leading to the minimum cost paths computation. Since Earth's surface is an ellipsoid, the planners require special handling to compute heuristics and edge costs between cities with higher precision, which is provided by a geodesic algorithm [5], [6].

## 2.2. Geodesic Algorithm

The algorithm for computation of geodesic distances [5], [6] considers geodetic latitudes and ellipsoidal longitudes, applied to an oblate ellipsoid [11]. An auxiliary sphere is created with corresponding variables, such as reduced latitudes and spherical longitudes [12]. It serves as a mathematical tool for the association between geographic coordinates, azimuths, flattening, and eccentricity. The result is a set of integrals expanded by Taylor Series [5], [13], and [14], giving final values for the azimuths, longitudes, and geodesic distances [6].

It works as follows: geographic coordinates of starting ($\phi_1, \lambda_1^0$) and ending vertices ($\phi_2, \lambda_2^0$) of an edge are represented in a canonical form [7]. These vertices have their reduced latitudes ($\beta_1, \beta_2$) computed. Earth's first eccentricity [13], [14] is employed at $\beta_1$ and $\beta_2$ calculations. Both vertices are also classified as antipodal or not, according to their geographic coordinates. Then, an initial value for the azimuth of the starting point $\alpha_1$ is estimated. In case of non-antipodal extremities, a small set of equations expressed at [15], [16], [12] and [6] gives the initial value of $\alpha_1$. Otherwise, specific procedures for antipodal points are described in [6]. In the current article, only non-antipodal extremities are considered.

The initial value of $\alpha_1$ must be further refined, leading to higher levels of precision. For this purpose, an auxiliary sphere is built to make an equivalence between the geodesic curve, which connects *start* vertex to *goal* vertex on an ellipsoid's surface, and an arch of a great circle. The values of $\alpha_1$, $\beta_1$, and $\beta_2$ are demanded for computation of spherical arc length variables ($\sigma_1, \sigma_2$). From $\sigma_1$ and $\sigma_2$, spherical longitudes ($\omega_1, \omega_2$) are obtained. From $\omega_1$ and $\omega_2$, the values of ellipsoidal longitudes ($\lambda_1, \lambda_2$) are given [17].

Then, the values for ellipsoidal longitudes *obtained* from the refinement procedure ($\lambda_1, \lambda_2$) are compared to the expected ones ($\lambda_1^0, \lambda_2^0$), defined earlier, in the beginning of the geodesic algorithm. The difference between the obtained and the expected values leads to an error, which is associated with the concept of reduced latitude $m_{12}$ [18] to give a new, and more precise, value for $\alpha_1$. If the error is not equal to zero or is higher than an acceptable bound, calculations of $\sigma_1, \sigma_2, \omega_1, \omega_2, \lambda_1, \lambda_2, m_{12}$ and $\alpha_1$ will be repeatedly performed (in a loop) until the ellipsoidal longitude error is reduced to a suitable value.

When the final value of $\alpha_1$ is obtained, the other variables used in the refinement process are updated one last time. They will be used in distance and longitude integrals [6].

The distance integral employs spherical arc length $\sigma$, an expansion factor for integrals $k$ and the polar semi-axis $b$. Factor $k$ can be obtained from Earth's second eccentricity $e'$ and the azimuth ($\alpha_0$) of the intersection point of the geodesic curve with the Equator [6].

Equation (1) shows the calculation of factor $k$.

$$k = e'.\cos(\alpha_0) \quad (1)$$

Equation (2) illustrates the distance integral.

$$\frac{s}{b} = \int_0^\sigma \sqrt{1 + k^2 sen^2(\sigma)} d\sigma \quad (2)$$

Equation (3) gives the longitude integral. It employs spherical arc length $\sigma$, factor $k$, Earth's first flattening $f$, spherical longitude $\omega$, and the azimuth $\alpha_0$.

$$\lambda = \omega - f sen(\alpha_0) \int_0^\sigma \frac{(2-f)}{(1-f)\sqrt{1 + k^2 sen^2(\sigma)} + 1} d\sigma \quad (3)$$

Equations (2) and (3) are expanded by Taylor Series up to $6^{th}$ order, which enable the integrals to be encoded as a finite sum of factors. The results of such sums are the values of ellipsoidal longitudes, which must be identical or very close to the values initially expected, and the length of the geodesic curve connecting two vertices in a graph.

The geodesic length is used by GPC as edge costs for graphs.

## III. SIMULATOR STRUCTURE

### 3.1. Databases

A graph $G = (V,E)$ is composed by a set $V$ of vertices and a set $E$ of edges. Each edge $e_{xy} \epsilon E$ connects elements $v_x \epsilon V$ and $v_y \epsilon V$. $G$ may be used for modelling aerial connections between cities. In this case, the cities are represented by the geographic coordinates of certain points located on Earth's surface. Such points are the elements of $V$. Connections between these points are elements of $E$.

The coordinates and connections for each city are registered in memory. Two text files store such data. The first file is reserved for the geographic coordinates of vertices, while the second one stores all direct connections from a certain city, i.e., the coordinates file stores the locations of $V$ elements, while the connections file keeps all $E$ members.

It is worth to mention that coordinates and connections files can be edited according to the user's needs, what modifies the format of the graph. The coordinates file can be edited when new cities must be inserted in the graph or when several cities, already inserted in the graph, must be removed from it.

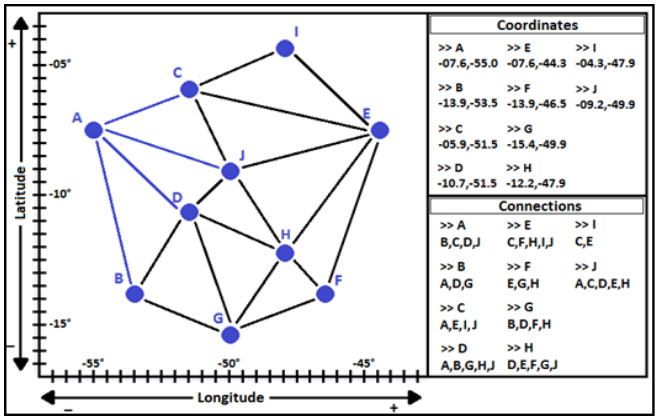

Figure 1.a. Position and connections of vertex A.

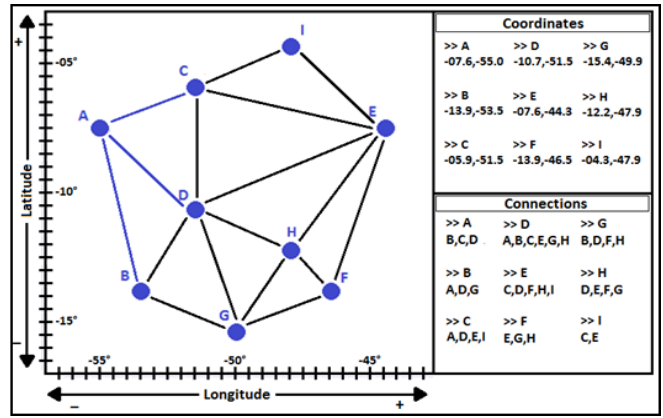

Figure 1.b. Connections after modifications.

The connections file can be edited whenever is necessary to modify direct connections from a city. The vertex degree may also be modified.

As an example shown in Fig. 1, one can consider a graph where a vertex *A*, represented by coordinates ($\phi_A, \lambda_A$), is connected to vertices *B*, *C*, *D* and *J*. Each city is represented by a set of coordinates and a set of connections. Vertices *E*, *F*, *G*, *H* and *I* are also present in the graph (Fig. 1.a). By removing vertex *J* and modifying the connection lists of vertices *A*, *C*, *D*, *E* and *H*, one creates another graph, with new edges between several vertices (Fig. 1.b). This graph modification can only be done by user's request.

In Fig. 1, the coordinates for each vertex are declared by its latitude and longitude, in this order. The values are given in decimal degrees. For simplification, the coordinates are represented with 1 decimal place of precision. However, in practical tests, the coordinates have 4 decimal places of precision. The connections for each vertex are represented by the adjacent cities, sorted in alphabetical order.

Both files (coordinates and connections) are read only once by a C program that creates an AVL tree, called *Original_AVL*, to store such data. The nodes of the tree represent cities locations on Earth's surface. Each node is identified by a name; a pair of geographic coordinates; an adjacencies list that registers all possible direct connections from a city; and pointers *father*, *left* and *right*, used in the assembly of the tree. Fig. 2 illustrates an example of the tree structure.

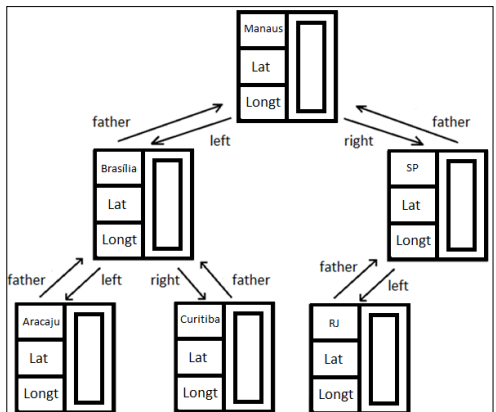

Figure 2. *Original_AVL* Structure.

### 3.2. GPC Structure

To calculate traversal costs between vertices, in kilometers, the planners need to consult coordinates stored in *Original_AVL* and run the geodesic algorithm. GPC supports A*, LPA* and D*Lite path-planning algorithms to find an optimal route between *start* and *goal* vertices and employs *Original_AVL* as a reference tree. The geodesic algorithm computes traversal costs and heuristics. The latter ones are geodesic distances between a certain vertex to start or goal destinations, depending of the planner executed.

GPC structure is shown in Fig. 3. The text files contain all information about vertices coordinates and connections, which must be kept in *Original_AVL*, the reference tree. The header file *AVL.h* is responsible for the creation and balance of *Original_AVL*, storing the content of the text files in the tree, as shown in Fig.2. The tree follows alphabetical in-order sort.

Another header file, named *Geodesy.h*, stores all functions necessary for the geodesic algorithm. The functions have, as input parameters, the geographic coordinates of two vertices. The result given by the geodesic algorithm is the length of the direct connection between the input vertices.

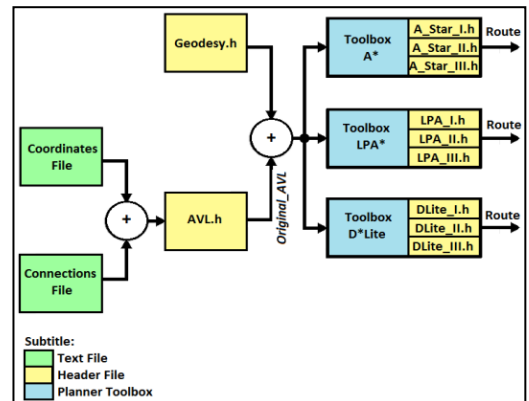

Figure 3. GPC Structure

### 3.3. Toolboxes

The planners must check the information stored in the reference tree and apply the geodesic functions in order to compute a final route, from a *start* to a *goal* vertex. To achieve this objective, each planning algorithm is implemented with a toolbox and three header files.

The toolbox is a set of data structures required for the execution of the planner. The header files contain functions which manage and modify the variables shown in the toolbox. The planner algorithm (GRPA) is written in one of such files and gives a route.

The toolboxes are used in this article as graphic representations of variables and structures. Each planner maintains the following structures inside a toolbox: a map of vertices, a priority queue, and a route. They also have pointers to handle special vertices, such as *start, goal* and, in case of D* Lite execution, *last* [8]. D* Lite needs an extra global variable, *km*, responsible for the correct calculation of new keys for the priority queue [8]. Fig 4 shows the structures and variables respectively related to A*, LPA*, and D* Lite.

According to Fig.4, the colored description indicates the exclusive variables for a certain planner, thus, it shows the small differences between toolboxes. The priority queue for A* algorithm is different from the queue for LPA* and D*Lite. Each toolbox has only one map of vertices, which can be *AVL_A*, AVL_LPA** or *AVL_D*Lite.* Two additional variables are used exclusively by D*Lite: pointer *last* and counter *km*.

Routes are linked lists whose elements are identified by a name, a pair of geographic coordinates, the final traverse cost $g$ for the respective element (optional), and a pointer to the next itinerary stop of the list.

Priority queues are linked lists, whose terms are identified by a name and composed by the variables employed in priority calculations and a pointer to the next priority of the queue. A* employs traversal cost $g$ and heuristic $h$ [1] to compute priorities, also known as *keys*, for the vertices. LPA* behaves similarly, using costs $g$ and $rhs$ and heuristic $h$ to create keys [2], which are expressed here by two terms: $K1$ and $K2$. D*Lite employs the same variables of LPA*, adding the extra global variable $km$ [3] to create keys.

The queues are sorted in ascending order, i.e. first element of the queue has the lowest key.

keys have the same $K1$, the tiebreaker is the value of $K2$, calculated by $min(g(s),rhs(s))$ [2]. Similarly, D*Lite generates $K1$ by $[min(g(s),rhs(s)) + h(start,s) + km]$, while $K2$ values are obtained by $min(g(s),rhs(s))$ [3].

A priority queue controls the way vertices will be analyzed by the respective planner during route computation. The procedure of scanning and upgrading a certain vertex is known as *expansion*. When a vertex is expanded, its traversal cost, heuristic and backpointer are updated. The adjacent vertices are inserted in the map (Fig.4) and also have heuristics and a traversal cost variable updated.

A backpointer, known as *back*, is the pointer that registers the adjacent vertex from which the traversal cost will be the lowest possible. Consequently, routes are always extracted by tracking backpointers.

Thus, priority queues are used by all planners to control vertices expansions while the algorithms are running. The vertex with the lowest key of the priority queue will be the first one to be expanded. The process of vertices expansions is carried out until a final route is obtained by the planner.

The maps of vertices are independent AVLs – *AVL_A*, AVL_LPA** and *AVL_D*Lite* – built by the planners from the information kept in *Original_AVL*. Each map is built and expanded only by request of its respective algorithm. Thus, *AVL_A** is built and expanded by A*, *AVL_LPA** is edited by LPA*, and *AVL_D*Lite* is controlled by D*Lite.

*Original_AVL* is not modified by the planners. However, the independent maps can be edited and built again from scratch. The structures of maps resemble the model for *Original_AVL*. However, there are some differences, since the nodes of maps have extra fields: a blockage status, mentioning the availability of a certain node to be used as an itinerary stop; three variables for traversal costs and heuristics ($f, g$ and $h$ for A*; $g, rhs$ and $h$ for LPA* and D*Lite); and a pointer for *back*. Fig. 5 illustrates a model for maps of vertices, showing that *AVL_A*, AVL_LPA** and *AVL_D*Lite* can be built in the same way.

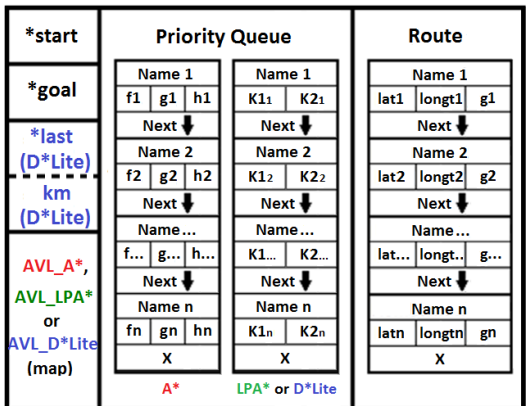

Figure 4. Toolboxes representation.

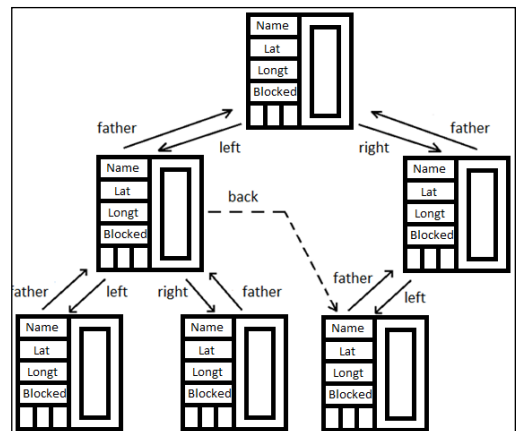

Figure 5. Example of a map of vertices – *AVL_A*, AVL_LPA** or *AVL_D*Lite*.

The A* queue is arranged in ascending order of $f$, where $f = g + h$. If two or more keys have the same $f$, the tiebreaker is the value of $g$ [1]. The LPA* queue is arranged in ascending order of $K1$, which is calculated by $min(g(s),rhs(s)) + h(s)$, where $s$ is the vertex associated to such key. If two or more

The only differences between the independent AVLs are the contents of their nodes, which are presented in Fig. 6. Colored variables are specific for a certain planner.

Every *AVL_A** node contains a name; a blockage status; an identifier of OPEN or CLOSED status [1]; a pair of coordinates; variables of costs and heuristics, useful for key calculations; pointers for tree structure (*father, left* and *right*); a backpointer and a list of adjacencies.

Similarly, *AVL_LPA\** nodes are identified by a name; a blockage status; a pair of coordinates; variables for priority computations; pointers for tree structure; backpointer and a list of adjacencies. The fields for *AVL_D\*Lite* nodes are the same as those for *AVL_LPA\**.

A* and LPA* expand vertices from *start* to *goal.* When they reach this objective, it is necessary to create a route from *start* to *goal,* a task performed by *back*. From *goal* vertex, the backtracking is done by successively following the backpointers from *goal* to *start*. Then, the final route is obtained by following the inverse path of the backtracking.

D*Lite expands vertices from *goal* to *start.* The backtracking is done from *start* to *goal* and the final route is the same path obtained by the backtracking search.

Each node of the independent AVLs has a list of adjacencies (Fig. 6). The lists of adjacencies found in the independent AVLs have elements composed by three fields: a name, a pointer to the next element and a value named *Dist*. The latter stores the geodesic distances between a certain city and its respective adjacency. *Dist* only appears in *AVL_A\**, *AVL_LPA\** and *AVL_D\*Lite* nodes.

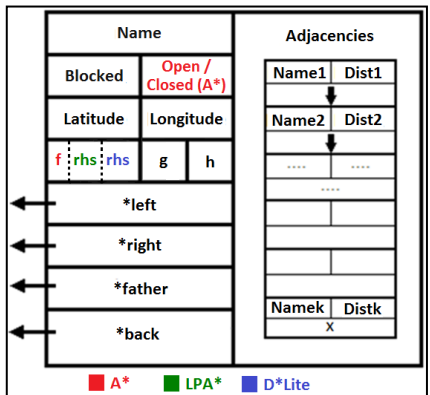

Figure 6. Internal fields for the nodes of a map of vertices.

An example with A* algorithm illustrates this: A city 'X' is directly connected to cities K, H and W. The city X has a node in *Original_AVL*, which is known as *X-node* of *Original_AVL*. Similarly, X has a node in *AVL_A\**, known as *X-node* of *AVL_A\**.

In the *X-node* of *Original_AVL,* the list is described as ({K}; {H}; {W}). However, in the *X-node* of *AVL_A\**, the list is described as ({K, $|\overline{KX}|$}; {H, $|\overline{HX}|$}; {W, $|\overline{WX}|$}).

Let *S* be the list of adjacencies found in the *X-node of AVL_A\**. Formally, *S* can be described as

$$S = \{(y_i, |\overline{xy_i}|)| \; 0 < i \leq n\} \quad (4)$$

where *x* is the city *represented* by the node that contains the list of adjacencies (X in the example) and $y_i$ is the set of all adjacent cities (K, H and W in the example) to *x*. The value *n* is the maximum value of adjacent vertices of *x*.

In other words, $x = X$, $y_1 = K$, $y_2 = H$, $y_3 = W$.

Thus, $S = (\{K, |\overline{KX}|\}; \{H, |\overline{HX}|\}; \{W, |\overline{WX}|\})$.

*Dist* values are memorized after the first expansion of X in *AVL_A\** because they are the *geodesic distances* used as traversal costs from X to all its neighbors. Such values are employed as edge costs while A* is computing a path. If they were not saved, they should always be calculated when A* needed some edge cost between two cities. The same logic applies to LPA* and D*Lite. This strategy aims to save time, avoiding repetitive executions of the geodesic algorithm.

To manage a toolbox, three header files were created for it. The first header file builds and balances the map of vertices. The second file controls the priority queue. The third file contains the planner code and employs the two previous files to generate the final route. In other words, there are three header files for each toolbox.

### 3.4. Planning algorithms implementations

As mentioned in the previous subsection, one of the header files designed for the planner contains the code lines and must deal with the data structures abstracted in the corresponding toolbox. The other two header files perform specific activities with some elements of the toolbox, like managing map of vertices and controlling priority queues.

For example, *DLite_III.h* is a header file that keeps all code lines for D*Lite execution. In the implementation proposed for GPC, the planner needs to manipulate a map of vertices (*AVL_D\*Lite*), a priority queue and some other variables abstracted inside a unit called *Toolbox D\*Lite* (Fig.3). With the purpose of dealing with *AVL_D\*Lite*, *DLite_I.h* was created. To work with the priority queue, *DLite_II.h* was written. GPC runs D*Lite by calling *DLite_III.h,* which contains the core of D*Lite and employs the other two headers, together with the toolbox, to calculate a final route.

The current session details how A* is described in *A_Star_III.h*, LPA* in *LPA_III.h* and D*Lite in *DLite_III.h*. In other words, the next paragraphs describe how the path-planning algorithms work in GPC.

A* [1] creates vertices *start* and *goal*, inserting them in the map. Keys are created for the vertices and inserted in the proper priority queue, considering that *g(goal)* is initially set to ∞. The map and the queue are initially empty. The status of *start* is set to *open*, and the vertex with the smallest key is selected for expansion. If the selected vertex for expansion is *goal*, one last update of neighbors is done, *goal* status is set to *closed* and the algorithm finishes. Otherwise, the status of the vertex is set to *closed*, its backpointer is updated, its neighbors will have their *f, g* and *h* values updated and the priority queue is modified to follow adjustments made in the map. Then, the algorithm selects the vertex with the smallest key for expansion in a loop procedure and checks, at each iteration, whether the vertex is *goal*.

LPA* [2] creates *start* vertex and its key, setting *g(start)* to ∞ and *rhs(start)* to zero. The key is inserted in the priority queue, which is initially empty. Then, *goal* vertex and its key are created, where *g(goal) = rhs(goal) =* ∞. While the lowest key of the queue is smaller than *goal*'s key or *rhs(goal)* differs from *g(goal)*, the vertex with the lowest key has its consistency checked and updated, its neighbors have their *rhs* values and *back* pointers updated, and the priority queue registers keys from inconsistent vertices. When edge costs are changed, the updates of consistency, *rhs* values and *back* pointers are performed again over the affected vertices, leading to the calculation of other routes. This algorithm is able to maintain maps and queues from successive searches for static starting and ending vertices.

D* Lite [3] initially sets *last = start*, *km = 0*, *g(goal) = ∞* and *rhs(goal) = 0*. The priority queue is initially empty. Then, *goal*'s key is inserted in the queue. *Start* vertex is created and *g(start) = rhs(start) = ∞*. While the lowest key of the queue is smaller than *start*'s key or *rhs(start)* differs from *g(start)*, the vertex with the lowest key will have its consistency checked and updated. Its neighbors will also have their *rhs* values and *back* pointers updated. This *while-loop* is executed until a route is calculated. Then, *start* vertex varies until it reaches *goal*. If obstacles are detected in the path, *km = km + h(last, start)*, *last = start* and the affected vertices are updated. The changing of km interferes directly in future computation of keys and is important to avoid unnecessary reordering of queues and to guarantee optimal paths. To compute a new path, the *while-loop* previously mentioned is executed again. D*Lite is able to maintain maps and queues for successive calculations with dynamic starting and static ending points.

Since the structures adopted by the planners are similar, the differences of performances are noticed when successive calculations are done after the change of blockage status of several nodes.

Cities are "blocked" when they are forbidden to be used as route stops. Their blockage status may be modified to force the planners to find alternative paths.

## IV. EXPERIMENTAL EVALUATION

The following section describes experiments where two cities are randomly chosen as starting and ending points. An initial path is computed. Then, several obstacles are randomly imposed in the path by changing blockage status of some intermediate nodes. The resulting routes must be the same for all three planning algorithms, but the time required by each of them varies according to the number of times a certain procedure must be done to achieve alternative paths. In the current session, three experiments are performed.

### 4.1. Experimental Objectives

The first experiment investigates numerical precision of edge costs calculated by GPC. Two vertices are randomly selected from *Original_AVL* and have a geodesic distance between them computed. In this experiment, the connections database is not yet employed. The only objective is to measure the length of a curve on an ellipsoid's surface and check its precision.

The second experiment compares time performances of A* and LPA* algorithms. Vertices *start* and *goal* are randomly selected from *Original_AVL* and the planners must compute a path from *start* to *goal*. Coordinates and connections databases are now employed, according to GPC structure (Fig.3). Blockage statuses of intermediate stops are randomly changed, leading to new routes. The length of routes and the time required to calculate them are registered for both algorithms and compared. *Start* and *goal* vertices do not change after successive searches.

The third experiment compares time performances of LPA* and D*Lite algorithms. Vertices *start* and *goal* are also randomly selected. Coordinates and connections databases are also employed. Both planners must compute a path from *start* to *goal*. Blockage statuses of intermediate stops are randomly changed. The length of new routes and the time required to calculate them are registered for both algorithms and compared. *Start* vertex changes after successive searches, while *goal* stays static.

### 4.2. Experimental Setup

All experiments require information provided by a Coordinate Database, which contains the latitudes and longitudes of 124 cities. Second and third experiments require Connections Database, which contains all possible direct connections from a city. Heuristics are consistent and the graph does not follow a geometric standard.

Simulations were performed in a machine with Fedora 23 64-bits OS, Intel Core 2 Duo T6600 (2.2Ghz) processor and 4GB of RAM.

The distances given by the execution of geodesic functions are compared to those measured on Google Earth, version 7.1.5. Google Earth is a freeware widely used to simulate routes between points located on Earth's surface. Among its tools, there is a GPS service, a flight simulator, a route calculator and a ruler to measure distances between points. The last two features are employed in this article.

Karney's algorithm and Google Earth [18] are both based on World Geodetic System (WGS) [19] [20], whose latest revision is WGS84 (last updated in 2004).

### 4.3. Experimental Results

**Experiment A:** The implementation of geodesic functions made for the present simulator had an average precision of 99,7%. This result was obtained after a set of tests, where geographic coordinates of two different points on Earth's surface were introduced as input parameters of the geodesic algorithm. The results were the geodesic lengths between both points. Then, such lengths were compared to distances measured on Google Earth and given by Karney's online simulator [8].

The best precision percentage was 99,9997%, while the lowest detected was 99,52%. Higher levels of imprecision may exist, but were not detected in a limited set of tests. Azimuths were approximately equal, proving that the orientation between points was correct.

Errors equal to or higher than 25 meters were rare. This result satisfies simulation demands, where the connections between cities are made from one airport to another. The average length of a typical runway is no less than 1.8 km. Therefore, the error levels do not put security at risk for the purposes of this simulator. The results were obtained with no software or hardware costs. Since the edge costs were properly computed, the next experiments can be performed with the necessary precision.

**Experiment B:** A graphic comparison of time efficiency between A* and LPA* algorithms is shown in Fig. 7. An arbitrary number of independent cycles are executed. In this case, 17 cycles are tested, since this quantity of cycles is sufficient for performance comparisons between planners. For each cycle, there is a static start, a static goal and ten iterations. For each iteration, the blockage status of an intermediate vertex is changed, forcing the creation of a new path, i.e., a cycle is completed after the computation of ten distinct routes for the same *start* and *goal* vertices. The execution time shown in the graph is the total sum of time after ten iterations of a cycle.

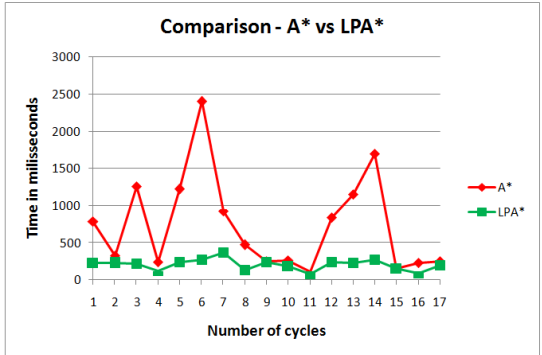

Figure 7.Time efficiency - A* vs LPA*.

For example, the first cycle simulates the execution of A* and LPA* for *start* city K, *goal* city L and ten different routes connecting K to L. The overall times for A* and LPA* are registered and printed in a graph. The second cycle works with *start* city H, *goal* city V and ten different routes connecting H to V. Overall times for A* and LPA* are collected again. Then, the next overall times for 15 different cycles are registered and illustrated (Fig. 7).

It is worth to mention that the times registered at one cycle do not interfere with the times collected for other cycles, meaning that each cycle is independent from the others.

The behaviors shown in the graph of Fig.7 can be explained by the LPA* ability to reuse information from previous searches, keeping maps of vertices and priority queues. This allows the algorithm to compute new routes with less effort, considering that edge costs are maintained in *AVL_LPA** nodes. Therefore, LPA* doesn't require frequent executions of the geodesic algorithm, saving time. A* needs to recalculate these values again and, depending of the vertices degrees, the computation of new routes can be more or less expensive in terms of time. LPA* is faster or, in the worst case, has the same performance than A*. The length of routes and the itinerary stops are the same for all situations, leading to optimized paths.

*AVL_A** and *AVL_LPA** nodes maintain edge costs between vertices in *Dist* field (Fig. 6). GPC uses this strategy to call the geodesic algorithm only once for an edge, reducing time consumption for both planners. GPC will only recalculate edge costs when a planning algorithm cannot maintain maps and priority queues.

Since each cycle has ten iterations, A* has to compute maps and queues from scratch ten times. LPA* can reuse such information, leading to different time performances between planners, as shown in Fig.7. The execution time for each cycle is the sum of time after ten iterations and is described in milliseconds. LPA* has a better performance and calculates the same routes faster.

In some tests, time difference between planners was small, typically lower than 200 milliseconds. In other tests, the difference is higher than 2 seconds. It is explained by the different degrees of vertices that are expanded during path computations, since higher degrees demands more geodesic calculations for edge costs.

**Experiment C:** A graphic comparison of time efficiency between LPA* and D*Lite is given by Fig. 8. Again, an arbitrary number of cycles is executed. In this case, there are 17 cycles. Each cycle is also complete after ten iterations of path computations and obstacle insertions. In Experiment C, however, there is a dynamic start, i.e., an intermediate vertex of the path will become the new *start* vertex at every iteration inside a cycle. According to LPA* and D*Lite definitions, LPA* needs to recalculate graph information considering the new *start* vertex. D*Lite can maintain it. Consequently, D*Lite will require less executions of the geodesic algorithm.

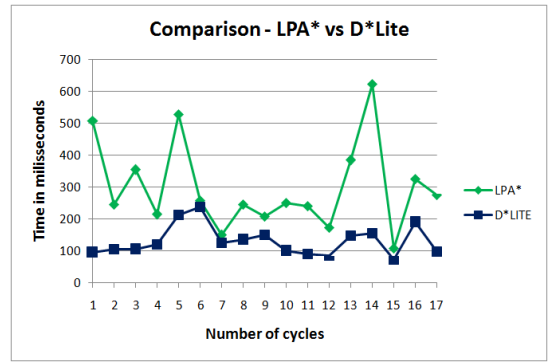

Figure 8.Time efficiency - LPA* vs D*Lite.

D*Lite saves information from past searches, using the map already created to compute a new route in less time. In most situations, the D*Lite performance is better or equal than LPA*'s in terms of time consumption.

In some cycles, the first iterations registered a faster performance for LPA*. It may be explained by the different direction of path calculation. Both planners must give a path from *start* to *end* vertices, but D*Lite calculates such path in a reverse order, from *end* to *start*. Thus the vertices initially expanded by LPA* are not the same as the ones expanded by D*Lite. If the vertices initially expanded by D*Lite have higher degrees than the ones processed by LPA*, the first iterations for D*Lite will take more time to be executed.

However, as the last iterations were run, the D*Lite ability to maintain its map and priority queue pays off, which surpass the initial disadvantage of expanding vertices with higher degrees. As a consequence, the total time after ten iterations, which is the execution time of the respective cycle, is lower for D*Lite. Such consequence is illustrated in Fig. 8 where, for all 17 different cycles, the total execution time is lower for D*Lite, proving its better performance.

This behavior can be noticed in bigger graphs. As D*Lite keeps more information, the difference of time performances becomes more visible with higher numbers of iterations at each cycle. As expected, the usage of consistent heuristics led to optimized paths. The length and itinerary stops are the same for both planners.

## V. RELATED WORK

Two other simulators are related to the GPC: online geodesic simulator [8] and the *D*Lite Demonstration* [9].

The online geodesic simulator [8] provides solutions for direct and inverse geodesic problems. GPC implementation of the geodesic algorithm solves a hybrid problem [6], largely based on the inverse geodesic problem. The online simulator was used during GPC development phase to validate its implementation of Karney's Algorithm and to check costs precision. The online geodesic simulator also offers tools to setup angular outputs and to compute geodesic lengths for different ellipsoids.

GPC calculates geodesic lengths in the form of Taylor Series, expanded up to $6^{th}$ order, and gives the final azimuths of both extremities of an edge. This method allows (2) and (3) to be described as a finite sum of terms in code lines, without affecting numerical precision. The order of expansion can be set arbitrarily, but the results expressed by Karney [6] show that a $6^{th}$ order expansion is sufficient to avoid the insertion of higher order terms.

*D*Lite Demonstration* [9] simulates a limited sized square-grid board, which operates as a graph. Each square of the board works as a vertex, while direct connections with neighboring squares represent edges. Such simulator allows path calculations between *start* and *end* squares, considering the number of neighbors, which can be set up to four or eight. Graphs can be generated manually or automatically. Among its settings, users are enabled to determine the density of blocked squares and the weight of unblocked ones, which affects traversal costs for each square.

GPC is partly based in such simulator [9], but is able to work with graphs of any size, where vertices dispositions don't follow a geometric standard. The degrees are not limited to a few options – vertices can be connected to a different number of adjacent nodes. GPC offers the execution of three different planners – D*Lite, LPA* and A* – while *D*Lite Demonstration* [9] works with one (D*Lite). GPC also offers a better edge cost precision by employing geodesy, which can be applied to real-life situations.

## VI. CONCLUSIONS

A novel GRPA simulator, named GPC, is described and evaluated. This simulator is adapted to perform the execution of three different planners: D*Lite, LPA*, and A*. GPC creates data structures to save databases and functions, allowing the executions of planners and a geodesic algorithm. Some experiments were performed, aiming the GPC evaluation and the comparison among those GPRA's. The experimental results show that GPC presents a precision of 99,7% in geodesic distance computations. Several advantages of GPC implementation can be mentioned, e.g., structs and data processing functions can be adapted to different planners; geodesic calculations are fast and precise; the maintenance of coordinates and connections is restricted to databases; and the reutilization of maps and priority queues reduces the number of geodesic algorithm executions.

Additionally, GPC is able to calculate distances with similar precision of the official online geodesic simulator and of widely known freeware (Google Earth). It also expands functionalities from the D*Lite demonstrator [9], since it compares three different path planners, employs consistent heuristics, works well on graphs with vertices located at non-standard positions and gives different route options automatically.

Future work includes the implementation of more recent planners, designed for situations where *goal* node is also dynamic (MTD*Lite and GAA*), and the adaptation of the current simulator for mobile robots, where sensors can detect changes of blockage status over time.